\documentclass[prb,twocolumn,superscriptaddress,longbibliography,notitlepage]{revtex4-1}
\usepackage{amssymb}
\usepackage{amsmath}
\usepackage{graphicx}
\usepackage{hyperref}
\usepackage{xcolor}
\usepackage{multirow}
\usepackage{amsfonts}
\usepackage[title]{appendix}

\newcommand{\comment}[1]{}

\begin{document}

\title{Investigating the Fermi-Hubbard model by the Tensor-Backflow method}

\author{Xiao Liang}
\email{xliang06@wm.edu}
\affiliation{Research Computing(Information Technology), College of William and Mary, Williamsburg, Virginia 23185, USA}

\begin{abstract}

	Recently, a variational wave-function based on the tensor representation of backflow corrections along with a Lanczos step has achieved state-of-the-art energy precision for Fermi-Hubbard-type models. However, solving the Fermi-Hubbard model remains challenging, and the validity of any method must be assessed through the ground state’s physical properties.
    For simplicity, we refer this method as Tensor-Backflow.
    In this work, we apply the Tensor-Backflow method to investigate the Fermi-Hubbard model on two-dimensional lattices up to 256 sites, exploring various interaction strengths $U$, electron fillings $n$, next-nearest-neighbor hopping $t'$, and boundary conditions. By considering backflow terms from nearest- or next-nearest-neighbor sites, we achieve competitive results without enforcing geometric symmetries on the variational wave-function.
	The optimizations were stable from a prior unrestrictied Hartree-Fock state, followed by adding backflow corrections. Meanwhile, changing interaction strengths in the prior unrestrictied Hartree-Fock state is helpful to bypass the local minima.
    When $t'$=0, by considering nearest-neighbor backflow terms, linear stripe order emerges successfully for the case of $n$=0.875 and $U$=8 on a $16 \times 16$ lattice with periodic boundary conditions. In a similar case with open boundary conditions, the energy obtained is only $4.5 \times 10^{-4}$ higher than the state-of-the-art method fPEPS with bond dimension $D$=20.
    Compared to state-of-the-art neural network methods, the energies obtained using the Tensor-Backflow approach are competitive, with relative errors below $5 \times 10^{-3}$.
    For $n$=0.8 and $n$=0.9375, direct optimizations yield results consistent with the phase diagram from AFQMC.
    When $t'$=-0.2, considering next-nearest-neighbor backflow terms leads to energies that are either competitive with or even lower than those from state-of-the-art neural network approaches. For instance, for $n$=0.875 and $U$=8 on a $12 \times 12$ lattice with periodic boundary conditions, the energy obtained is $8.1 \times 10^{-4}$ lower than that from the neural network result.
    Thus, the Tensor-Backflow method demonstrates strong representational capabilities for solving the Fermi-Hubbard model.

\end{abstract}
\maketitle

\section{Introduction}

The Fermi-Hubbard model is a fundamental quantum lattice model used to describe correlated electrons and plays a key role in understanding the mechanisms of superconductivity~\cite{Hubbard_SC1,Hubbard_SC2,Hubbard_SC3}.
However, solving the Fermi-Hubbard model, particularly on finite-sized lattices, presents significant challenges.
Near the ground state, the energies of states with different ordering can be extremely close, making the optimization process easy to getting trapped in local minima.
To mitigate boundary effects, the lattice size must be sufficiently large when studying finite systems.
As such, identifying the ground state properties requires numerical methods that are both unbiased and scalable.
To this end, various numerical approaches have been developed and benchmarked for solving the Fermi-Hubbard model~\cite{Hubbard_benchmark,Manybody_benchmark,Hubbard_DMET,Hubbard_DMRG,Hubbard_NN1,Hubbard_NN2,Hubbard_NN3,Hubbard_NN4,Hubbard_TNS,tJ_fPEPS,Hubbard_fPEPS}.

For instance, the density-matrix renormalization group (DMRG) is well-suited for one-dimensional or quasi-one-dimensional systems~\cite{Hubbard_benchmark,Hubbard_DMRG,Hubbard_DMRG2,Hubbard_DMRG3,Hubbard_DMRG4}, but its accuracy is not satisfactory for two-dimensional systems. The Quantum Monte Carlo (QMC) method offers high accuracy but suffers from high computational complexity, particularly due to the ``sign problem" in certain systems~\cite{QMC_sign_1,QMC_sign_2,QMC_sign_3}. The projected-entangled-pair state (PEPS) can achieve high accuracy for two-dimensional lattices after variational optimization~\cite{tJ_fPEPS,Hubbard_fPEPS,PEPS_liu}, though its computational cost remains high, especially for periodic boundary conditions (PBC). Similarly, both infinite-PEPS (iPEPS)~\cite{iPEPS} and density-matrix embedding theory (DMET)~\cite{Hubbard_DMET} impose constraints on the ground state due to the size of the supercell used.

The variational Monte Carlo (VMC) method avoids the ``sign problem", but its effectiveness depends critically on the choice of variational wave-function. Recently, neural-network-based variational wave-functions have been extensively benchmarked~\cite{RBM_science,CNN_J1J2_1,CNN_J1J2_2,CNN_J1J2_3,J1J2_SOTA1,J1J2_SOTA2,J1J2_SOTA3,Hubbard_NN1,Hubbard_NN2}. Due to their deep structure, neural networks can capture long-range correlations and entanglements in quantum states, enabling state representations that go beyond the area-law entanglement~\cite{NN_represent1,NN_represent2}.
While backflow corrections implemented via neural networks have demonstrated high accuracy in continuous systems~\cite{NN_continuous1,NN_continuous2,NN_continuous3}, their application to the lattice Fermi-Hubbard model has challenges. The high optimization difficulties of neural networks lead to unsatisfactory energy precision for large lattices~\cite{Hubbard_NN1,Hubbard_NN2} or require specific preconditions on the wave-function~\cite{Hubbard_NN3,Hubbard_NN4}.
Recently, the Tensor-Backflow along with a Lanczos step has achieved state-of-the-art for Fermi-Hubbard-type models~\cite{tensorbackflow}. This approach combines backflow corrections~\cite{backflow1,backflow2} with a tensor representation of the wave-function. The optimization of the Tensor-Backflow wave-function is stable from optimizing a prior unrestrictied Hartree-Fock (UHF), followed by including backflow corrections in the wave-function.

In this work, we apply the well-established Tensor-Backflow method~\cite{tensorbackflow} to investigate the Fermi-Hubbard model across a range of electron fillings $n$, interaction strengths $U$, next-nearest-neighbor hopping strengths $t'$, and boundary conditions on lattices as large as 256 sites. Without enforcing geometric symmetries, the achieved energies are competitive with, or even exceed, those of current state-of-the-art methods.
For $t'$=0, we obtain a ground state with linear stripe order for $n$=0.875 and $U$=8 on a $16 \times 16$ lattice under periodic boundary conditions (PBC), without applying any pinning fields. The periods of spin-density-wave (SDW) and charge-density-wave (CDW) are 16 and 8, respectively. For higher interaction strengths, $U$=10 and $U$=12, the linear stripe order is preserved when starting from the wave-function converged for $U$=8, and the energies are lower than those for other orders obtained through direct optimization.
For $n$=0.8 with $U$=4, 8 and $n$=0.9375 with $U$=4, the results are consistent with the phase diagram from AFQMC~\cite{AFQMC_phase_diagram,AFQMC_n0.9375}. In more challenging cases with $t'$=-0.2, where non-zero superconductivity is expected~\cite{Hubbard_SC1}, the energies achieved by Tensor-Backflow is competitive with, or even lower than, those from state-of-the-art neural network states (NQS)~\cite{Hubbard_NN3}, demonstrating the strong representational power of the Tensor-Backflow approach.

This paper is organized as follows:
Sec.~(\ref{sec:tensor-backflow method}) introduces the Tensor-Backflow method, including the variational Monte Carlo (VMC) optimization and Lanczos step, and discusses the computational complexity of the method.
Subsec.~(\ref{subsec:optimization details}) compares two optimization strategies.
Subsec.~(\ref{subsec:energy comparisons}) presents energy comparisons with current state-of-the-art methods, including fPEPS, AFQMC, NQS, and others.
Subsec.~(\ref{subsec:n0.875}) presents results for $n$=0.875 with both $t'$=0 and $t'$=-0.2, including energy comparisons, ground state patterns, and charge and spin structure factors.
Subsec.~(\ref{subsec:n0.8 and 0.9375}) discusses results for $n$=0.8 and $n$=0.9375, providing energy results as well as charge and spin structure factors.
Sec.~(\ref{sec:conclusions}) concludes the paper.

\section{the Tensor-Backflow method}
\label{sec:tensor-backflow method}
\subsection{Variational wave-function definitions}
In this subsection, we introduce the Tensor-Backflow method developed in the previous literature~\cite{tensorbackflow}.
The backflow correction is originally derived from the transformation of the position of particles~\cite{backflow1}:
\begin{equation}
    \mathbf{r}_\alpha^B=\mathbf{r}_\alpha+\sum_\beta\eta_{\alpha\beta}[\mathbf{S}](\mathbf{r}_\beta-\mathbf{r}_\alpha),
    \label{r_transformation}
\end{equation}
where $\mathbf{r}_\alpha$ are actual particle positions and $\eta_{\alpha\beta}[\mathbf{S}]$ are variational parameters depending on the many-body state $|\mathbf{S}\rangle$.
The original backflow correction is constructed by a linear combination of orbitals~\cite{backflow1,backflow2}:
\begin{equation}
    \phi^B_{k\sigma_k}(\mathbf{r}_{i,\sigma_i})=\phi_{k\sigma_k}(\mathbf{r}_{i,\sigma_i})+\sum_{j}c_{ij}[\mathbf{S}]\phi_{k\sigma_k}(\mathbf{r}_{j,\sigma_j}),
    \label{orbital_transformation}
\end{equation}
where $c_{ij}[\mathbf{S}]$ is the coefficient depending on the configuration $|\mathbf{S}\rangle$ and $\phi_{k\sigma_k}$ is the Hatree-Fock (HF) orbital.
In the original backflow~\cite{backflow1,backflow2}, the coefficients $c_{ij}$ in Eq.(\ref{orbital_transformation}) are assumed to depend on local configurations $\mathbf{s}(\mathbf{r}_i)$ and $\mathbf{s}(\mathbf{r}_j)$, rather than on the full state $|\mathbf{S}\rangle$ for simplicity. The final wave-function is the Slater determinant: $w(\mathbf{S})=\mathrm{det}M_{ik}^B$, where $M_{ik}^B=\langle \mathbf{r}_i|\phi^B_{k\sigma_k}\rangle$.

The backflow corrections defined by Eq.(\ref{r_transformation}) and (\ref{orbital_transformation}) are applied only to positions.
As a result, the matrix $M_{ik}^B$ in the Slater determinant is quasi-diagonal with respect to different spins, which limits the representation ability for Hamiltonians with spin couplings. To enhance the representation ability for spin-coupled systems, we consider an unrestricted form of backflow corrections.
This is achieved by introducing backflow terms that depend not only on positions but also on spins. The orbital is constructed as~\cite{tensorbackflow}:
\begin{equation}
	\phi^B_{k\sigma_k}(\mathbf{r}_{i,\sigma_i})=\phi_{k\sigma_k}(\mathbf{r}_{i,\sigma_i})+\sum_{j}c_{ij}\sum_{\sigma_j=\pm 1}\phi_{k\sigma_k}(\mathbf{r}_{j,\sigma_j}).
	\label{BW_unrestricted_orbital}
\end{equation}
Here in Eq.(\ref{BW_unrestricted_orbital}), we adopt the similar assumption for the coefficient $c_{ij}$ as in the original backflow in Eq.(\ref{orbital_transformation}).

For the state representation, the backflow corrected orbital can involve complex relationships among the orbital index $k\sigma_k$ and the position $\mathbf{r}_j$ in the HF orbital and the configurations of $\mathbf{s}(\mathbf{r}_i)$ and $\mathbf{s}(\mathbf{r}_j)$ in $c_{ij}$. This complexity goes beyond the representation ability of the product of $c_{ij}$ and $\phi_{k\sigma_k}$~\cite{Hubbard_NN1,Hubbard_NN3}.
To achieve the enhanced representation ability based on the variables in Eq.(\ref{BW_unrestricted_orbital}), we develop a mathematical form capable of representing arbitrary functions of these variables. A straightforward approach is to represent Eq.(\ref{BW_unrestricted_orbital}) by a single high-dimensional tensor: $g_{\mathbf{r}_i,k\sigma_k,j,\mathbf{s}_i,\mathbf{s}_j}$, where each index in the tensor corresponds to an independent variable in Eq.(\ref{BW_unrestricted_orbital})~\cite{tensorbackflow}.
This construction ensures high representation ability for backflow corrections on lattices.

Here we construct the tensor $g$, as well as the final wave-function, by considering the independent variables in Eq.(\ref{BW_unrestricted_orbital}).
The variables include the particle positions $\mathbf{r}_i$, the orbital index $k\sigma_k$ with $\sigma_k = \pm 1$, the backflow indices $j$, and the configurations $\mathbf{s}(\mathbf{r}_i)$ and $\mathbf{s}(\mathbf{r}_j)$.
Based on the definition, the final dimension of $g$ is~\cite{tensorbackflow}:
\begin{equation}
	[M,N,Q,d,d],
	\label{g_dimension}
\end{equation}
where $M$ denotes the number of sites, $N$ is the total number of particles, and $d$ represents the number of degrees of freedom at each site.
The first and second occurrences of $d$ correspond to $\mathbf{s}(\mathbf{r}_i)$ and $\mathbf{s}(\mathbf{r}_j)$, respectively.
In this work, we consider backflow contributions from both the site $\mathbf{r}_i$ and its surrounding sites.
Consequently, $Q$=5 when backflow terms from nearest-neighbor sites are included, and $Q$=$M$ when backflow terms from all sites are considered.

The final wave-function is a Slater determinant of fictitious particles:
\begin{equation}
w(\mathbf{S})=\mathrm{det}M^B,
	\label{ws}
\end{equation}
where the matrix element $M_{ik}^B$ is constructed based on the tensor representation of Eq.(\ref{BW_unrestricted_orbital}):
\begin{equation}
	M_{ik}^B=\sum_{j}g_{\mathbf{r}_i,k\sigma_k,j,\mathbf{s}(\mathbf{r}_i),\mathbf{s}(\mathbf{r}_j)},
	\label{M_ik}
\end{equation}
where $j$ is the summation index including the backflow terms.

Because $g$ is a single tensor, correlations between $\mathbf{s}(\mathbf{r}_i)$ and $\mathbf{s}(\mathbf{r}_j)$ are achieved naturally by the backflow rule. In contrast, in the literature~\cite{CP_tensor}, such correlations are introduced through products of decomposed tensors.
Compared with tensor decompositions, the cost for using a single tensor is the large parameter number. The total number of variational parameters equals the product of the dimensions of all indices of $g$. However, only a subset of these parameters is involved in generating a $w(\mathbf{S})$.

By constructing $g$, each site is associated with a high-dimensional tensor whose indices include $k\sigma_k$ and $\mathbf{s}(\mathbf{r}_i)$, as well as those of its surroundings sites $j$ and $\mathbf{s}(\mathbf{r}_j)$. Although the procedure for generating $w(\mathbf{S})$ is very different, the similar structure of $g$ to that of a two-dimensional tensor network results in high efficiency for first-order optimization~\cite{tJ_fPEPS, PEPS_liu}.
In the construction of the wave-function, no constraints are imposed on translational or rotational symmetries. The flexibility of the representation ability of $g$, combined with the straightforward construction of the wave-function, leads to the effectiveness of first-order gradient methods.

To stabilize the initial optimization and reduce the computational cost, we start by optimizing a prior UHF state. The UHF is represented by a tensor: $g_{\mathbf{r}_i,k\sigma_k,\sigma_i}^{\mathrm{UHF}}$, where $\sigma_i$ denotes the spin of the particle at site $\mathbf{r}_i$. Based on the UHF, we extend the tensor by introducing additional indices to incorporate backflow corrections. The resulting tensor takes the form: $g_{\mathbf{r}_i,k\sigma_k,\sigma_i,2,j,\mathbf{s}(\mathbf{r}_j)}$~\cite{tensorbackflow}, where the dimension of 2 distinguishes the double occupation at site $\mathbf{r}_i$.
Although the energy convergence of an UHF is very stable, the use of a prior UHF may introduce bias into the variational wave function~\cite{UHF}.
In this work, we avoid such bias when generating the results in Sec.~(\ref{numerical}) and compare the prior UHF with the converged wave-function in App.(\ref{app:compare_UHF_backflow}).

\subsection{Optimization method}
We use the VMC method to optimize the wave-function first, followed by a Lanczos step.
In the VMC, the energy and the gradient of the $\alpha$-th parameter are estimated through the Markov-Chain-Monte-Carlo (MCMC) process~\cite{Hubbard_fPEPS, PEPS_liu}:
\begin{equation}
	\begin{split}
	&E=\langle E_{\mathrm{loc}}\rangle\\
	&G^\alpha= 2\langle E_{\mathrm{loc}}O_{\mathrm{loc}}^\alpha \rangle
	-2\langle E_{\mathrm{loc}} \rangle\langle O_{\mathrm{loc}}^\alpha \rangle,
		\label{E and G}
	\end{split}
\end{equation}
where the local energy is $E_\mathrm{loc}(\mathbf{S})=\sum_{\mathbf{S}'}\frac{w(\mathbf{S}')}{w(\mathbf{S})}\langle \mathbf{S}'| \hat{H}|\mathbf{S} \rangle$, the $O_{\mathrm{loc}}(\mathbf{S})^\alpha=\frac{1}{w(\mathbf{S})}\frac{\partial w(\mathbf{S})}{\partial\alpha}$, and $\langle \cdots \rangle$ denotes the average on MC samples.

In the VMC optimization, the variational parameters are updated using the gradient descent method. Since the absolute values of the gradients are difficult to be estimated with finite MCMC samples, we update the parameters based on the sign of the gradient and apply a constant step size $\delta$:
$\alpha'=\alpha-\delta\mathrm{sgn}(G^\alpha)$. This parameter update scheme has been successfully applied in optimizing high-dimensional tensors, such as the PEPS~\cite{tJ_fPEPS, PEPS_liu}.

To extend the representation ability beyond backflow corrections, the wave-function is further optimized using a Lanczos step.
In the Lanczos procedure, the wave-function is optimized by constructing Krylov subspaces:
\begin{equation}
	|\Psi_{p=1}\rangle=A|\Psi_{p=0}\rangle+B|\Psi_{p=0}^{\perp}\rangle,
	\label{lanczos}
\end{equation}
where $A$ and $B$ are parameters to be determined, and $|\Psi_{p=0}\rangle$ is the wave-function optimized by the VMC. The orthogonal wave-function is built by $|\Psi_{p}^\perp\rangle=(\hat{H}-E_p)/\sigma_p|\Psi_{p}\rangle$, where the energy expectation and the variance are: $E_{p}=\langle \Psi_{p}|\hat{H}|\Psi_{p} \rangle,\ \sigma_{p}^2=\langle\Psi_{p}| (\hat{H}-E_p)^2|\Psi_{p} \rangle$.

\subsection{Computational complexity}
For the Tensor-Backflow wave-function, the complexity of generating a coefficient $w(\mathbf{S})$ is $\mathcal{O}(N^3 Q)$, as derived from Eq.(\ref{ws}) and Eq.(\ref{M_ik}), where $\mathcal{O}(N^3)$ comes from the Slater determinant and $\mathcal{O}(Q)$ from the summation in Eq.(\ref{M_ik}).

When considering backflow terms from nearest-neighbor (NN) sites, $Q$=5. The representation ability is improved by including backflow terms from next-nearest-neighbor (NNN) sites, where $Q$=9. We refer to these two cases as NN and NNN backflows, and the computational complexity is $\mathcal{O}(N^3)$ for both types. The maximal representation ability is achieved by including backflow terms from all sites, where $Q$=$M$, leading to a complexity of $\mathcal{O}(M N^3)$.

For generating a coefficient after Lanczos step, the complexity is $\mathcal{O}(N^4 Q)$, based on Eq.(\ref{lanczos}). During the Lanczos step, the term $\langle \Psi_{p=0} | \hat{H}^2 | \Psi_{p=0} \rangle$ contributes to the calculation of the variance $\sigma_p^2$, resulting in a complexity of $\mathcal{O}(N^5 Q)$.

\section{Numerical results}
\label{numerical}
In this section, we present numerical results for the two-dimensional Fermi-Hubbard model achieved by the Tensor-Backflow method.
To demonstrate the method's high accuracy, we firstly present energy comparisons respect to state-of-the-art results.

The Hamiltinian of the Fermi-Hubbard model is:
\begin{equation}
	\begin{split}
	\hat{H}=&-t\sum_{\langle ij\rangle,\sigma}(\hat{c}_{i\sigma}^\dagger\hat{c}_{j\sigma}+\mathrm{H.c.})-t'\sum_{\langle\langle ij\rangle\rangle,\sigma}(\hat{c}_{i\sigma}^\dagger\hat{c}_{j\sigma}+\mathrm{H.c.})\\
	&+U\sum_i \hat{n}_{i\uparrow}\hat{n}_{i\downarrow},
	\end{split}
	\label{H_Hubbard}
\end{equation}
where $t$ and $t'$ are strengths for NN and NNN hoppings, respectively. $U$ is the strength of on-site interactions, $\hat{c}_{i\sigma}^\dagger(\hat{c}_{i\sigma})$ creates (destroys) a particle of spin $\sigma$ on the $i$-th site, and the particle number operator $\hat{n}_{i\sigma}=\hat{c}_{i\sigma}^\dagger\hat{c}_{i\sigma}$.
We set $t$=1 through our investigations.
Energy results achieved by wave-functions before and after a Lanczos step are evaluated with the MC sample number in the magnitude of $\mathcal{O}(4.5\times 10^4)$.

\begin{figure}[t]
\includegraphics[width=0.95\columnwidth]{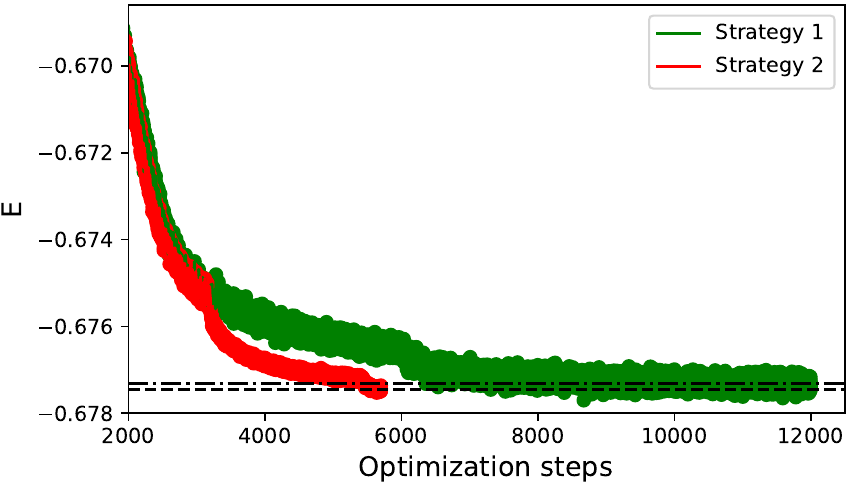}
\caption{Comparisons of two optimization strategies demonstrated by $n$=0.875, $U$=8 and $t'$=0 on the $4\times 16$ lattice under OBC, with NN backflow in wave-functions. Both strategies start from the same initial state. For strategy 1, sample number is fixed as $\mathcal{O}(4.5\times 10^4)$, the step size is reduced from $2\times 10^{-3}$ to $2\times 10^{-4}$ at the 6000-th step. For strategy 2, sample number is increased from $\mathcal{O}(4.5\times 10^4)$ to $\mathcal{O}(4.5\times 10^5)$ at around 3000-th step, meanwhile maintaining the step size to $2\times 10^{-3}$. After about another 2500 steps, the step size is reduced to $2\times 10^{-4}$. Strategy 1 and 2 converge to -0.6773 and -0.6774, respectively.}
\label{fig:convergence_compare}
\end{figure}

\subsection{Optimization details}
\label{subsec:optimization details}
In VMC optimizations, the number of samples and the step size are crucial for achieving full convergence.
To determine the optimal values for these parameters, we tested two optimization strategies:
\begin{itemize}
\item[(1)] With a sample size of $\mathcal{O}(4.5 \times 10^4)$, the step size is initially set to $2 \times 10^{-3}$ and then reduced to $2 \times 10^{-4}$.
\item[(2)] With a sample size of $\mathcal{O}(4.5 \times 10^4)$, the step size is initially set to $2 \times 10^{-3}$, the sample size is then increased to $\mathcal{O}(4.5 \times 10^5)$ while maintaining the step size, and finally, the step size is reduced to $2 \times 10^{-4}$.
\end{itemize}
The comparison between these two strategies is shown in Fig.(\ref{fig:convergence_compare}). The specific case depicted in the figure is $n$=0.875, $U$=8, $t'$=0 on a $4 \times 16$ lattice under open boundary conditions (OBC), with NN backflow terms in the wave-function. From the figure, both strategies converge to similar energies. However, strategy (2) requires fewer optimization steps. Both strategies yield similar ground state patterns after full convergence, and further comparisons of the ground state patterns are provided in App.~(\ref{app:optimization_compare}).
Additionally, we find that strategy (2) is more effective for the convergence of all-site backflow due to the significantly larger number of MCMC samples. Therefore, strategy (2) is used for optimizing wave-functions with all-site backflow. For lattices larger than 64 sites, both strategies require no more than $2 \times 10^4$ optimization steps to achieve full convergence.

\begin{figure}[t]
\includegraphics[width=\columnwidth]{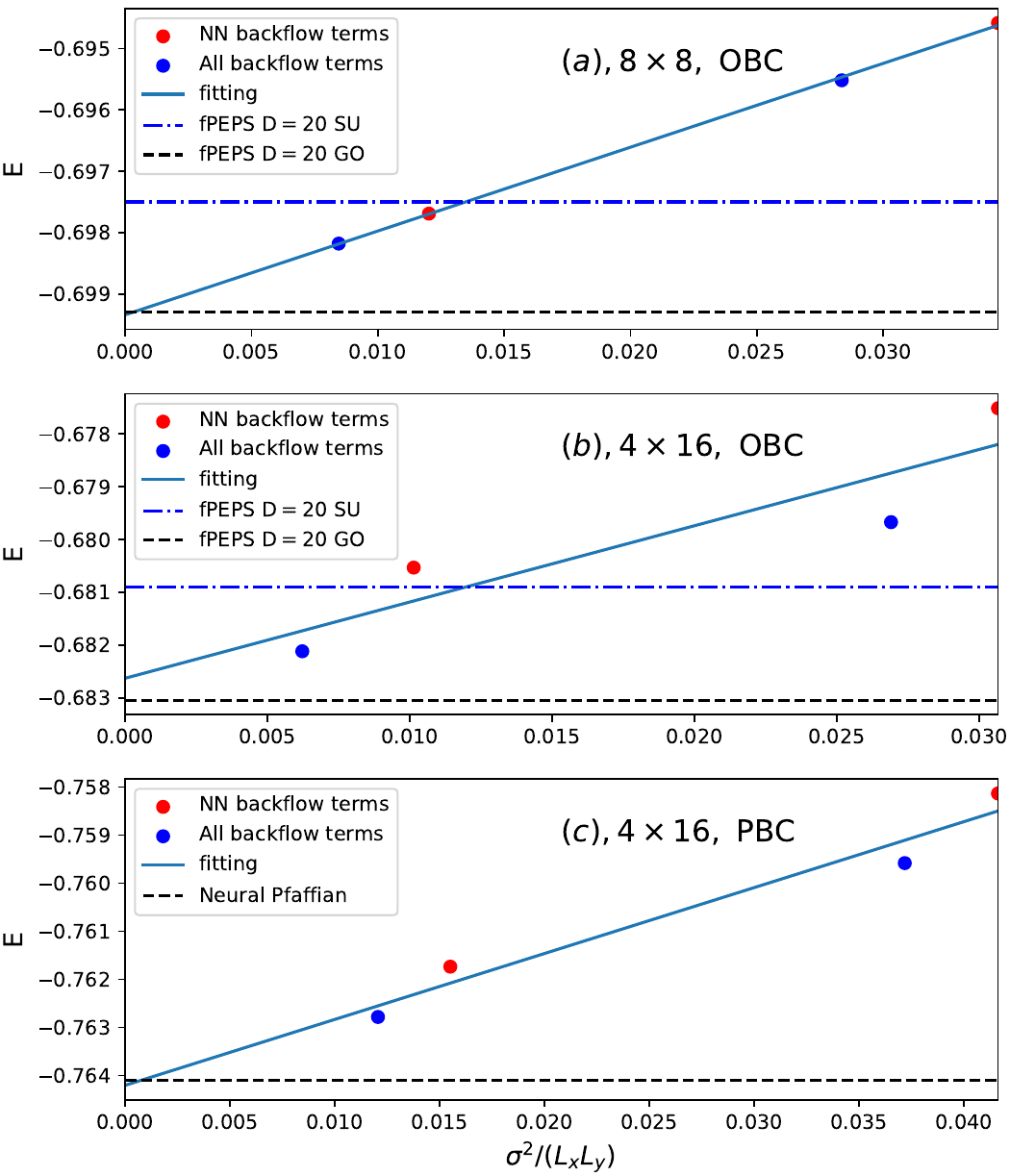}
\caption{Energy extrapolations and comparisons respect to fPEPS~\cite{Hubbard_fPEPS} under OBC for the lattice of $8\times 8$(a) and $4\times 16$(b). All cases are for $n$=0.875, $U$=8 and $t'$=0. Energy extrapolation for the lattice of $4\times 16$ under PBC(c) yields to the neural Pfaffian result~\cite{Hubbard_NN4}. $\sigma^2$ is the energy variance. Red and blue solid dots depict Tensor-Backflow with NN and all-site backflow terms. For solid dots of the same color, low and high energies depict results with and without one Lanczos step. }
\label{fig:E_compare}
\end{figure}
\subsection{Comparisons with previous results}
\label{subsec:energy comparisons}
Fig.(\ref{fig:E_compare})(a)(b) depict energy comparisons respect to the fPEPS method under OBC on 64-site lattices.
The fPEPS is a state-of-the-art method for lattice fermions and its state representation ability can be improved by increasing the bond-dimension ($D$).
From literature~\cite{Hubbard_fPEPS}, the complexity of fPEPS scales rapidly respect to $D$: $\mathcal{O}(D^6)$, and the maximal bond dimension is $D$=20.
The fPEPS is firstly optimized by the imaginary-time-evolution method named as the simple-update (SU), followed by the gradient-optimization (GO). The fPEPS with $D$=20 can achieve very high energy precision under OBC.


In Fig.(\ref{fig:E_compare})(a)(b), red and blue solid dots represent the results for NN and all-site backflow terms, respectively. The higher and lower energy values for each color correspond to results without and with one Lanczos step, respectively. This demonstrates that a single Lanczos step significantly improves the energy precision for both NN and all-site backflow terms.
When comparing the energies obtained after one Lanczos step to those from the fPEPS with bond dimension $D$=20, we find that the NN backflow achieves energy precision comparable to the SU-optimized fPEPS, while the all-site backflow achieves precision competitive with the GO-optimized fPEPS. The Tensor-Backflow method imposes no limitations on boundary conditions, and energies for the $4 \times 16$ lattice under PBC are shown in Fig.(\ref{fig:E_compare})(c).

\begin{table}[]
\begin{tabular}{|c|c|c|c|cc|c|}
\hline
\ \ $n$\ \      & \ $U$\       & \ \ \begin{tabular}[c]{@{}c@{}}Lattice \\ Size\end{tabular}\ \    & \begin{tabular}[c]{@{}c@{}}Boundary \\ Condition\end{tabular}   & \multicolumn{1}{c|}{\ \ $E_{p=0}$\ \ \ } & \ \ \ $E_{p=1}$\ \  & \ \ \ Ref.\ \ \  \\ \hline
\multirow{5}{*}{0.875} & \multirow{4}{*}{8} & $8\times 8$ & \multirow{3}{*}{OBC} & -0.6955 & -0.6982 & -0.6993~\cite{Hubbard_fPEPS} \\ \cline{3-3}
 &  & $4\times 16$ &  & -0.6797 & -0.6821 & -0.6830~\cite{Hubbard_fPEPS} \\ \cline{3-3}
 &  & $16\times 16$ &  & -0.7219 & -0.7257 & -0.7260~\cite{Hubbard_fPEPS} \\ \cline{3-4}
 &  & $16\times 16$ & PBC & -0.7509 & -0.7552 & -0.7544~\cite{Manybody_benchmark} \\ \cline{2-4}
 & 10 & $16\times 16$ & A\textendash PBC & -0.6786 & -0.6828 & -0.6813~\cite{Hubbard_TNS} \\ \cline{1-4}
\multirow{2}{*}{1} & \multirow{2}{*}{8} & \multirow{2}{*}{$12\times 12$} & PBC & -0.5209 & -0.5224 & -0.5246~\cite{AFQMC_n1} \\ \cline{4-4}
 &  &  & A\textendash PBC & -0.5205 & -0.5224 & -0.5249~\cite{AFQMC_n1} \\ \cline{1-4}
44/64 & \multirow{3}{*}{4} & $8\times 8$ & \multirow{3}{*}{PBC} & -1.1833 & -1.1846 & -1.1867~\cite{AFQMC_sym} \\ \cline{1-1} \cline{3-3}
80/100 &  & $10\times 10$ &  & -1.1086 & -1.1109 & -1.1147~\cite{AFQMC_sym} \\ \cline{1-1} \cline{3-3}
116/144 &  & $12\times 12$ &  & -1.1038 & -1.1061 & -1.1102~\cite{AFQMC_sym} \\ \hline
\end{tabular}
\caption{Energy comparisons respect to state-of-the-art results for various filling number $n$ and interaction strength $U$. All cases are for $t'$=0. A-PBC denotes the boundary condition of antiperiodic along the x-boundary and periodic along the y-boundary. $E_{p=0}$ and $E_{p=1}$ are energies achieved by Tensor-Backflow without and with one Lanczos step, respectively. }
	\label{tab:E_compare}
\end{table}

To perform energy extrapolation under the limit of zero energy variance, we linearly fit the four solid dots for each lattice size~\cite{Hubbard_TNS, lanczos}. The extrapolated energies are indicated by blue solid lines in Fig.(\ref{fig:E_compare}). From Fig.(\ref{fig:E_compare})(a)(b), the extrapolated energies under OBC are -0.6993 for the $8 \times 8$ lattice and $-0.6826$ for the $4 \times 16$ lattice. These extrapolated values are in close agreement with state-of-the-art energies: -0.6993 for the $8 \times 8$ lattice and $-0.6830$ for the $4 \times 16$ lattice, as achieved by the fPEPS with $D$=20 and GO optimization.
The extrapolated energy for PBC shown in Fig.(\ref{fig:E_compare})(c) is $-0.7642$, which matches the state-of-the-art neural Pfaffian result~\cite{Hubbard_NN4}.

Energy comparisons with fPEPS~\cite{Hubbard_fPEPS}, NQS~\cite{Manybody_benchmark}, tensor networks~\cite{Hubbard_TNS}, and AFQMC~\cite{AFQMC_n1,AFQMC_sym} are provided in Tab.(\ref{tab:E_compare}). In the table, $E_{p=0}$ represents the energy from the Tensor-Backflow wave-function, and $E_{p=1}$ represents the energy obtained after a Lanczos step on the converged wave-function. For OBC, the reference energies are taken from GO-optimized fPEPS with $D$=20~\cite{Hubbard_fPEPS}. For the $8 \times 8$ and $4 \times 16$ lattices under OBC, the wave-function corresponds to all-site backflow, while for the $16 \times 16$ lattice under OBC, the wave-function uses NN backflow.

In this work, the relative error of energy $E$ is defined as: $(E-E_\mathrm{ref})/|E_\mathrm{ref}|$.
where $E_{\mathrm{ref}}$ is the reference energy.
For OBC, the relative errors of $E_{p=1}$ with respect to the fPEPS results are $1.6 \times 10^{-3}$, $1.3 \times 10^{-3}$, and $4.1 \times 10^{-4}$ for the $8 \times 8$, $4 \times 16$, and $16 \times 16$ lattices, respectively. This demonstrates that Tensor-Backflow achieves high precision for OBC.
For other boundary conditions, the reference energy from the VMC method in the literature~\cite{Manybody_benchmark} is $E_{\mathrm{ref}}$=-0.7544 for $n$=0.875 and $U$=8 on the $16 \times 16$ lattice under PBC, with $E_{p=1}$ being $1 \times 10^{-3}$ lower. For $n$=0.875 and $U$=10, under antiperiodic along the x-boundary and periodic along the y-boundary, denoted by A-PBC in Tab.(\ref{tab:E_compare}), the reference energy is obtained from energy extrapolation results of the tensor network method in the literature~\cite{Hubbard_TNS}, with $E_{p=1}$ being $2.2 \times 10^{-3}$ lower.
In Tab.(\ref{tab:E_compare}), for each reference energy from the literature~\cite{AFQMC_n1, AFQMC_sym}, the energy after one Lanczos step  is competitive, with relative errors on the order of $\mathcal{O}(10^{-3})$.

\begin{table}[]
\begin{tabular}{|c|c|c|c|cc|c|}
\hline
\ \ $n/U$\ \  &
  \begin{tabular}[c]{@{}c@{}}Boundary\\ Condition\end{tabular} &
  \ $t'$\  &
  \ \begin{tabular}[c]{@{}c@{}}Lattice\\ Size\end{tabular}\  &
  \ $E_{p=1}$\ \ \  &
  \ \ Ref.~\cite{Hubbard_NN3}\  &
  \ \begin{tabular}[c]{@{}c@{}}Relative\\ Error\end{tabular}\  \\ \hline
\multirow{5}{*}{0.875/8} & \multirow{4}{*}{PBC} & \multirow{2}{*}{0}    & $8\times 16$  & -0.7568 & -0.7568 & --                  \\ \cline{4-4}
                         &                      &                       & $16\times 16$ & -0.7552 & -0.7563 & $1.5\times 10^{-3}$ \\ \cline{3-4}
                         &                      & \multirow{2}{*}{-0.2} & $8\times 16$  & -0.7410 & -0.7419 & $1.2\times 10^{-3}$ \\ \cline{4-4}
                         &                      &                       & $12\times12$  & -0.7420 & -0.7414 & -$8.1\times 10^{-4}$ \\ \cline{2-4}
                         & OBC                  & 0                     & $16\times 16$ & -0.7257 & -0.7275 & $2.5\times 10^{-3}$ \\ \hline
\end{tabular}
    \caption{Energy comparisons respect to state-of-the-art NQS results~\cite{Hubbard_NN3} for $n$=0.875, $U$=8 under various $t'$, lattice sizes and boundary conditions, with NN backflow for $t'$=0 and NNN backflow for $t'$=-0.2. The relative error is defined as: $(E_{p=1}-E_\mathrm{ref})/|E_\mathrm{ref}|$. $E_{p=1}$ are competitive or even lower than state-of-the-art NQS results. In particular, The result for $t'$=-0.2 on the $12\times 12$ lattice under PBC is lower than the NQS result.}
	\label{tab:E_compare_with_NQS}
\end{table}
\begin{table}[]
\begin{tabular}{|c|c|c|c|c|}
\hline
\ \begin{tabular}[c]{@{}c@{}}Backflow\\ Type\end{tabular}\  &
  \ \begin{tabular}[c]{@{}c@{}}Lattice\\ Size\end{tabular}\  &
  \ \begin{tabular}[c]{@{}c@{}}Parameter\\ Number\end{tabular}\  &
  \ \ \begin{tabular}[c]{@{}c@{}}Relative\\ Error~\cite{Hubbard_fPEPS}\end{tabular}\ \ \  &
  \ \ \begin{tabular}[c]{@{}c@{}}Relative\\ Error~\cite{Hubbard_NN3}\end{tabular}\ \ \  \\ \hline
\multirow{2}{*}{NN}       & $8\times 8$   & 286720  & $2.3\times 10^{-3}$ & $2.6\times 10^{-3}$ \\ \cline{2-2}
                          & $4\times 16$  & 286720  & $3.6\times 10^{-3}$ & $4.1\times 10^{-3}$ \\ \cline{1-2}
\multirow{2}{*}{all-site} & $8\times 8$   & 3670016 & $1.6\times 10^{-3}$ & $1.9\times 10^{-3}$ \\ \cline{2-2}
                          & $4\times 16$  & 3670016 & $1.3\times 10^{-3}$ & $1.8\times 10^{-3}$ \\ \cline{1-2}
NN                        & $16\times 16$ & 4587520 & $4.5\times 10^{-4}$ & $2.5\times 10^{-3}$ \\ \hline
\end{tabular}
	\caption{The parameter number used in Tensor-Backflow for various site number and backflow types, and comparisons of energy precision respect to state-of-the-art fPEPS~\cite{Hubbard_fPEPS} and NQS~\cite{Hubbard_NN3}, with $n$=0.875 and $U$=8 under OBC. Relative errors are for $E_{p=1}$. Parameter number is based on Eq.(\ref{g_dimension}), where $Q=5,M$ for NN and all-site backflows, respectively. }
	\label{tab:parameter_number_compare}
\end{table}

From the comparisons presented in Tab.(\ref{tab:E_compare}), the Tensor-Backflow method achieves results that are competitive with those from mainstream approaches. Furthermore, we compare our results to the recent state-of-the-art NQS method~\cite{Hubbard_NN3} in Tab.(\ref{tab:E_compare_with_NQS}).
In the table, all results (except those for OBC, detailed in Sec.~(\ref{subsec:n0.875})) were obtained using first-order gradient optimization, with fewer than $2 \times 10^{4}$ optimization steps and without applying any pinning fields during the optimization process. In contrast, for the NQS method, each optimization involved an improved stochastic reconfiguration procedure, pre-training, the application of pinning fields, and a large optimization step number~\cite{Hubbard_NN3}.
In particular, for the case with $t'$=-0.2, $n$=0.875, $U$=8 on the $12 \times 12$ lattice under PBC, the achieved energy from the Tensor-Backflow method is $8.1 \times 10^{-4}$ lower than the corresponding NQS result. Moreover, this optimization was performed using the first-order gradient method in fewer than $1 \times 10^{4}$ steps, with the optimization procedure detailed in Sec.~(\ref{subsec:n0.875}).
Therefore, the Tensor-Backflow method not only demonstrates strong representational capabilities but is also efficient in terms of optimization.

Tab.(\ref{tab:parameter_number_compare}) shows the number of parameters used in the Tensor-Backflow method for various lattice sizes and backflow types, along with the corresponding energy precision compared to two state-of-the-art methods under OBC.
According to Eq.(\ref{g_dimension}), for a filling of $n$=0.875 on 64-site lattices, the parameter numbers for NN and all-site backflow are 286,720 and 3,670,016, respectively. For the $16 \times 16$ lattice with NN backflow, the parameter number is 4,587,520.
In the table, we compare the energy obtained after one Lanczos step ($E_{p=1}$) with results from the state-of-the-art fPEPS~\cite{Hubbard_fPEPS} and NQS~\cite{Hubbard_NN3} methods. From the energy comparisons, we observe that NN backflow achieves relative energy errors on the order of $\mathcal{O}(10^{-3})$. Moreover, incorporating all backflow terms further improves the energy precision.

\begin{figure}[t]
\includegraphics[width=\columnwidth]{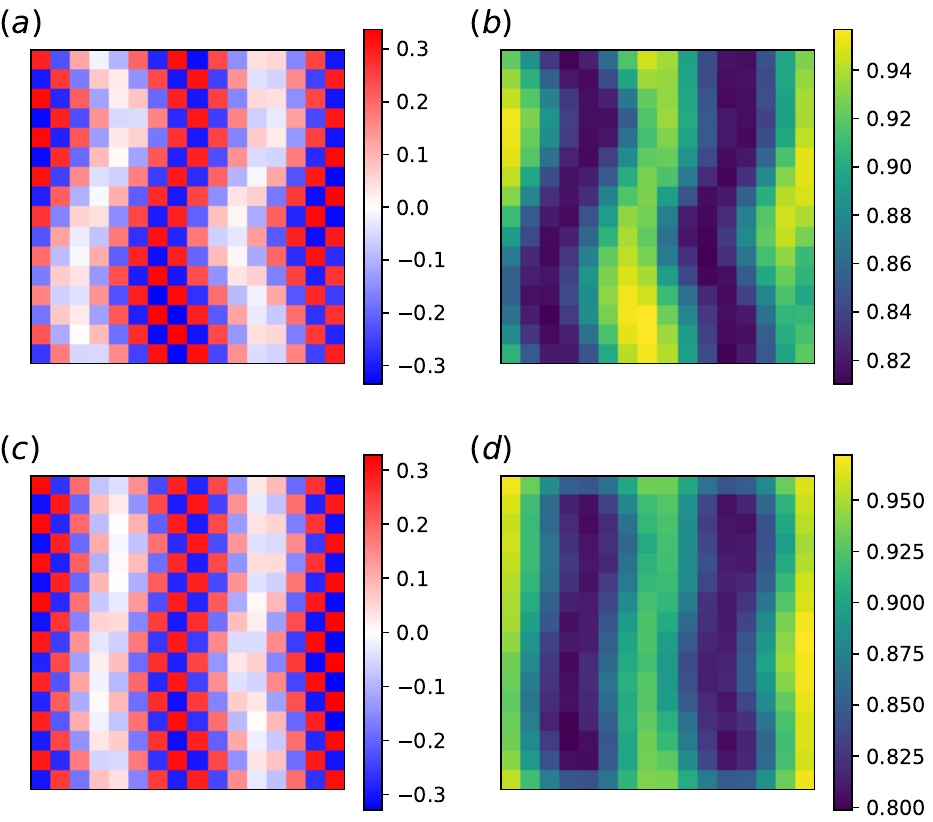}
\caption{Patterns of ground states for $n$=0.875, $U$=8 and $t'$=0 on the $16\times 16$ lattice under PBC(a)(b) and OBC(c)(d). All results are from $p$=1 wave-functions with NN backflow. Left-sided figures denote the average spin per site: $s_i^z=(n_i^\uparrow-n_i^\downarrow)/2$, and right-sided figures denote the average density per site: $n_i=n_i^\uparrow+n_i^\downarrow$. }
\label{fig:Sz_ni}
\end{figure}

The number of parameters in Tensor-Backflow is significantly lower compared to the state-of-the-art fPEPS method~\cite{Hubbard_fPEPS}. For the $16 \times 16$ lattice under OBC, the fPEPS with bond dimension $D$=14 contains $5.7 \times 10^6$ parameters, achieving an energy of $-0.7224$. The fPEPS with $D$=16 has $9.4 \times 10^6$ parameters and achieves an energy of $-0.7251$, while the fPEPS with $D$=18 uses $1.5 \times 10^7$ parameters and yields an energy of $-0.7260$. The fPEPS with the maximum bond dimension, $D$=20, requires $2.2 \times 10^7$ parameters and produces an energy similar to that obtained with $D$=18~\cite{Hubbard_fPEPS}.
For Tensor-Backflow on the $16 \times 16$ lattice, the energy without a Lanczos step ($E_{p=0}$) is $-0.7219$, and after the Lanczos step ($E_{p=1}$), it improves to $-0.7257$. While achieving comparable energy precision, Tensor-Backflow requires fewer parameters than fPEPS with $D$=14 when no Lanczos step is applied. After the Lanczos step, Tensor-Backflow achieves the same energy precision with significantly fewer parameters than fPEPS with $D$=18 or $D$=20. This demonstrates that Tensor-Backflow is a highly efficient state representation method.

\subsection{n=0.875}
\label{subsec:n0.875}
\begin{table}[]
\begin{tabular}{|c|c|c|c|c|}
\hline
\ \begin{tabular}[c]{@{}c@{}}Lattice\\ Size\end{tabular}\  &
\ \begin{tabular}[c]{@{}c@{}}Boundary \\ Condition\end{tabular}\  &
\ \ $U$\ \ &
\ \begin{tabular}[c]{@{}c@{}}Direct\\ Optimization\end{tabular}\  &
\ \begin{tabular}[c]{@{}c@{}}Transfer\\ from $U$=8\end{tabular}\ \\ \hline
\multirow{4}{*}{$16\times 16$} & \multirow{3}{*}{PBC} & 8  & -0.7552 & --      \\ \cline{3-3}
                       &                      & 10 & -0.6815 & -0.6831 \\ \cline{3-3}
                       &                      & 12 & -0.6303 & -0.6317 \\ \cline{2-3}
                       & A-PBC                & 10 & -0.6813 & -0.6828 \\ \hline
\end{tabular}
	\caption{Energy comparisons of $E_{p=1}$ between different initializations for $n$=0.875 and $t'$=0, with NN backflow in wave-functions. For the direct optimization of each case, the initial UHF is insufficiently optimized with $U_\mathrm{UHF}=U$. The direct optimization of $U$=8 achieves the ground state of linear stripe order, and such order is preserved when it is transferred for other $U$ and boundary conditions. Energy convergences of transferring from $U$=8 are detailed in App.(\ref{app:transfer_convergence}). Energies with linear stripe orders are much lower than those by direct optimizations.}
	\label{tab:lstripe}
\end{table}

The filling of $n$=0.875 is important for understanding the mechanisms of superconductivity, particularly when the NNN hopping term $t'$ is non-zero~\cite{Hubbard_SC1}. The ground-state energies at this filling exhibit strong competition among various methods~\cite{Hubbard_stripe}. In this subsection, we provide a detailed analysis of the ground-state properties and structure factors for $n$=0.875, considering both $t'$=0 and $t'$=-0.2. We begin by presenting the results obtained from the Tensor-Backflow method for the case of $t'$=0.

In this work, most of the results obtained from Tensor-Backflow are achieved through direct optimization. For a direct optimization, a UHF state is initially optimized, followed by the construction of the Tensor-Backflow by expanding the tensor's dimensions to include backflow terms. The Tensor-Backflow wave-function is then optimized using one of the strategies outlined in Sec.~(\ref{subsec:optimization details}), with no pinning fields applied during the optimization process.
When optimizing the UHF state, the interaction strength is denoted as $U_\mathrm{UHF}$. Unless otherwise specified, we set $U_\mathrm{UHF} = U$. It is important to note that the UHF is not fully optimized when $U_\mathrm{UHF} = U$, in order to avoid introducing prior bias~\cite{UHF}. Furthermore, Tensor-Backflow can override the properties of the initial UHF state, meaning that the converged wave-function is, in essence, a different state from the original UHF. A detailed comparison of the UHF and Tensor-Backflow states is provided in App.(~\ref{app:compare_UHF_backflow}).

Fig.(\ref{fig:Sz_ni})(a) and (b) show the linear stripe order of the ground state for $U$=8 on a $16 \times 16$ lattice under PBC, obtained through direct optimization. In Fig.(\ref{fig:Sz_ni})(a), we present the average spin per site, defined as $s_i^z = (n_i^\uparrow - n_i^\downarrow) / 2$, while Fig.(\ref{fig:Sz_ni})(b) shows the average density per site, $n_i = n_i^\uparrow + n_i^\downarrow$.
From the ground-state patterns, we observe that the periods for the spin-density wave (SDW) and charge-density wave (CDW) are 16 and 8, respectively.
For PBC, the ground state should be translational invariant thus an obvious pattern is not consistent with the true ground state.
Since translational symmetry is not enforced in the variational wave-function, the optimization may converge to a subspace of the true ground state, even though the obtained energy is competitive.

For $n$=0.875, $U$=8, $t'$=0 on a $16 \times 16$ lattice with OBC, we use the wave-function converged under PBC as the initial state and perform further optimizations. The achieved state-of-the-art energy is -0.7257, as reported in Tab.(\ref{tab:E_compare}) and (\ref{tab:parameter_number_compare}). Meanwhile, Fig.(\ref{fig:Sz_ni})(c) and (d) show the average spin and charge density per site, respectively.
Although the stripe order in the initial state has the same period as the ground state under OBC, the optimization process slightly modifies the stripe pattern to accommodate the boundary conditions. The resulting ground-state patterns under OBC are consistent with those from the fPEPS method~\cite{Hubbard_fPEPS}.

For other interaction strengths such as $U$=10 and $U$=12, Tab.(\ref{tab:lstripe}) presents energy comparisons between the linear stripe order and the results from direct optimizations. The linear stripe order is obtained by transferring the wave-function converged for $U$=8 and PBC. Notably, for $U$=12 under PBC, the direct optimization leads to a diagonal stripe order. Ground-state patterns from direct optimizations are shown in App.(~\ref{app:compare_UHF_backflow}), and the energy convergence of transferring from $U$=8 is detailed in App.(~\ref{app:transfer_convergence}).
Energy comparisons are summarized in Tab.(\ref{tab:lstripe}), with all energies evaluated using the $p$=1 wave-function. Since the energy of the linear stripe order is significantly lower than that of other orders achieved by direct optimizations, we conclude that the ground state for large $U$ is the linear stripe order.

\begin{figure}[t]
\includegraphics[width=\columnwidth]{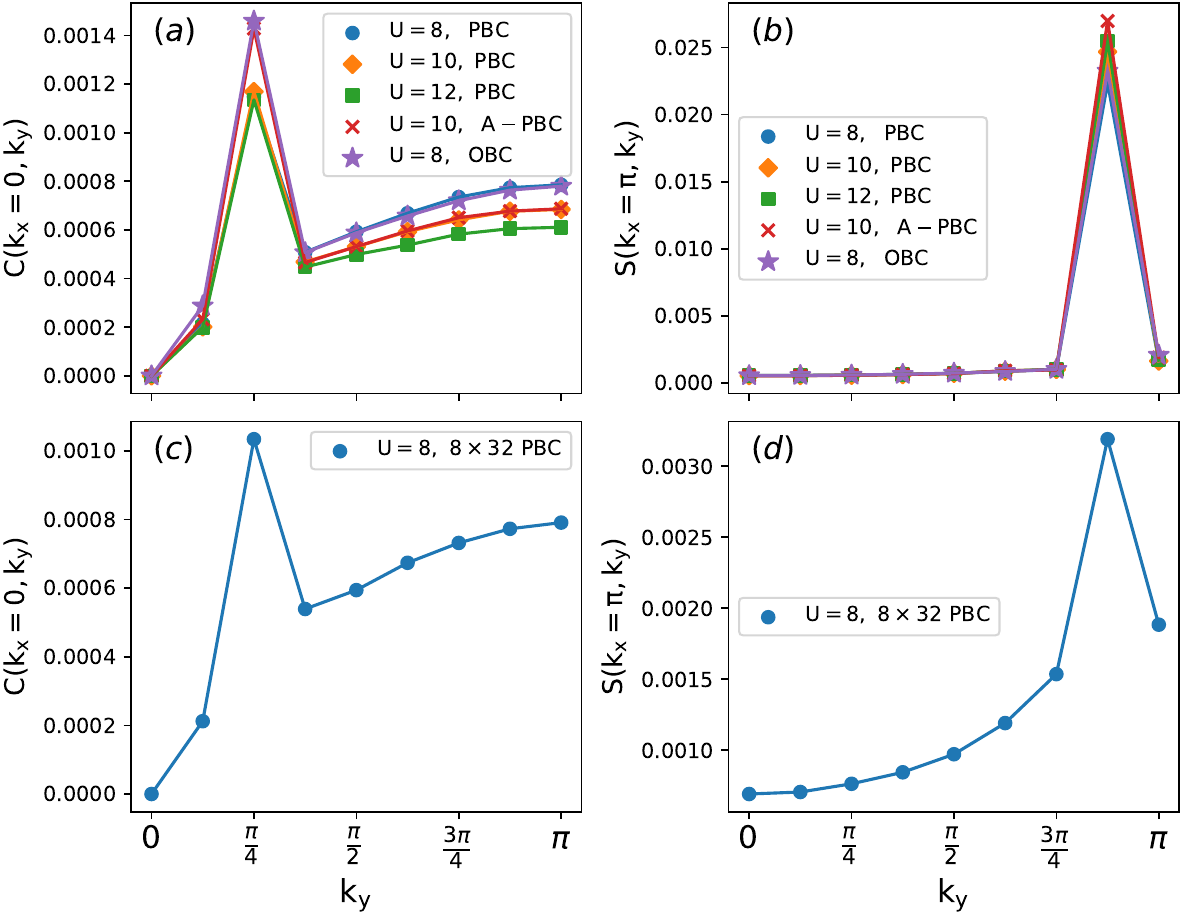}
\caption{Charge and spin structure factors for $n$=0.875 and $t'$=0 under various $U$ and boundary conditions. All results are from $p$=1 wave-functions, with NN backflow in wave-functions. (a)(b)On the $16\times 16$ lattice, results of $U$=8 under PBC are obtained by a direct optimization, meanwhile other results are obtained by the initialization with the $U$=8 PBC wave-function. (c)(d)Results of $U$=8 under PBC by a direct optimization on the $8\times 32$ lattice. }
\label{fig:Sk_Ck}
\end{figure}

\begin{table}[]
\begin{tabular}{|c|c|c|cc|}
\hline
\ \ \begin{tabular}[c]{@{}c@{}}Lattice\\ Size\end{tabular}\ \  & \ \ \ $t'$\ \ \  & \ \begin{tabular}[c]{@{}c@{}}Backflow\\ Type\end{tabular}\  & \multicolumn{1}{c|}{\ \ \ $E_{p=0}$\ \ \ } & \ \ \ $E_{p=1}$\ \ \  \\ \hline
$8\times 16$  & \multirow{3}{*}{0}    & \multirow{4}{*}{NN}  & -0.7529 & -0.7568 \\ \cline{1-1}
$8\times 24$  &                       &                      & -0.7500 & -0.7546 \\ \cline{1-1}
$8\times 32$  &                       &                      & -0.7505 & -0.7547 \\ \cline{1-2}
$8\times 16$  & \multirow{4}{*}{-0.2} &                      & -0.7354 & -0.7406 \\ \cline{1-1} \cline{3-3}
$8\times 16$  &                       & \multirow{3}{*}{NNN} & -0.7359 & -0.7410 \\ \cline{1-1}
$12\times 12$ &                       &                      & -0.7335 & -0.7388 \\ \cline{1-1}
$16\times 16$ &                       &                      & -0.7335 & -0.7386 \\ \hline
\end{tabular}
	\caption{Energy results obtained by direct optimizations for $n$=0.875 and $U$=8. For all cases in the table, initial UHFs are insuffieintly optimized with $U_\mathrm{UHF}=U$. For $t'$=-0.2, NNN backflow terms are considered in the Tensor-Backflow. On the $8\times 16$ lattice, NNN backflow achieves the energy slightly lower than that comparing to NN backflow. }
	\label{tab:additional_energies}
\end{table}
\begin{table}[]
\begin{tabular}{|c|c|c|cc|}
\hline
$n, U, t'$ & \begin{tabular}[c]{@{}c@{}}Backflow\\ Type\end{tabular} & \ \ $U_\mathrm{UHF}$\ \  & \multicolumn{1}{c|}{\ \ $E_{p=0}$\ \ } & \ \ $E_{p=1}$\ \  \\ \hline
\multirow{3}{*}{0.875, 8, -0.2} & \multirow{3}{*}{NNN} & 0 & -0.7344 & -0.7395 \\ \cline{3-3}
 &  & 4 & -0.7370 & -0.7420 \\ \cline{3-3}
 &  & 8 & -0.7335 & -0.7388 \\ \hline
\end{tabular}
    \caption{Comparisons of converged energies on the $12\times 12$ lattice from different $U_\mathrm{UHF}$, $U_\mathrm{UHF}$ is the interaction strength when optimizing the UHF. The UHF is fully converged when $U_\mathrm{UHF}\neq U$, otherwise the UHF is insufficiently optimized when $U_\mathrm{UHF}=U$. Initialization with $U_\mathrm{UHF}$=4 gives the lowest energy result for Tensor-Backflow. }
    \label{tab:compare_initial_UHF}
\end{table}

Structure factors provide insight into the periods of stripe order in the ground state. The charge and spin structure factors are defined as: $C(\textbf{k})=\frac{1}{M^2}\sum_{ij}e^{i\textbf{k}\cdot (\textbf{r}_i-\textbf{r}_j)}n_in_j$,
$S(\textbf{k})=\frac{1}{M^2}\sum_{ij}e^{i\textbf{k}\cdot (\textbf{r}_i-\textbf{r}_j)}s_i^zs_j^z$,
where $n_i$ and $s_i^z$ are defined in Fig.(\ref{fig:Sz_ni}). Note that $C(\mathbf{k})$ is set to zero when $k_x=k_y=0$. All results for structure factors are computed using the $p$=1 wave-functions.
Fig.(\ref{fig:Sk_Ck}) shows the structure factors for a filling of $n$=0.875 at various interaction strengths $U$ and boundary conditions, on lattices of $16 \times 16$ and $8 \times 32$. As seen in Fig.(\ref{fig:Sk_Ck})(a) and (b), the charge structure factor peaks at $\pi/4$, and the spin structure factor peaks at $7\pi/8$, corresponding to periods of 8 and 16 for charge-density-wave (CDW) and spin-density-wave (SDW) orders, respectively.
Fig.(\ref{fig:Sk_Ck})(c) and (d) present results from a direct optimization on the $8 \times 32$ lattice for $n$=0.875 and $U$=8 under PBC. A vertical stripe order is observed, with structure factor peaks at the same locations as those on the $16 \times 16$ lattice. On the $16 \times 16$ lattice, the side length of 16 corresponds to a full period of the SDW, which results in higher peak values compared to those on the $8 \times 32$ lattice.

Tab.(\ref{tab:additional_energies}) presents additional energy results for $n$=0.875 and $U$=8 under PBC for both $t'$=0 and $t'$=-0.2. All values in the table are obtained through direct optimizations. For $t'$=0 on the $8 \times 16$ lattice, a vertical stripe order emerges, with the same periodicity observed on the $16 \times 16$ lattice.

Energy results for $t'$=-0.2 are shown in Tab.(\ref{tab:additional_energies}), which plays a crucial role in understanding the mechanisms of superconductivity~\cite{Hubbard_SC1}. From the comparisons presented in the table, we observe that NNN backflow ($Q$=9) yields slightly lower energies than NN backflow ($Q$=5) for the $t'$=-0.2 case on the $8 \times 16$ lattice. Due to the non-zero NNN hopping strengths, NNN backflow is considered for all calculations involving $t'$=-0.2.

\begin{figure}[t]
    \includegraphics[width=\linewidth]{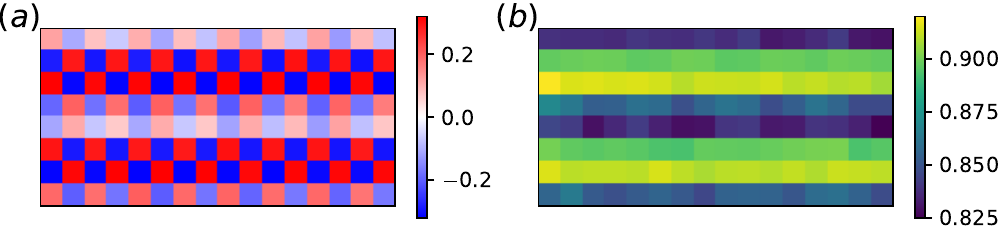}
    \caption{Patterns of ground states for $n$=0.875, $U$=8 and $t'$=-0.2 under PBC on the $8\times 16$ lattice, from the $p$=1 wave-function with NNN backflow. Left-sided (right-sided) figure denotes the average spin (density) per site, within definitions from Fig.(\ref{fig:Sz_ni}). }
    \label{fig:Sz_ni_NNN}
\end{figure}
\begin{figure}[t]
    \includegraphics[width=0.8\linewidth]{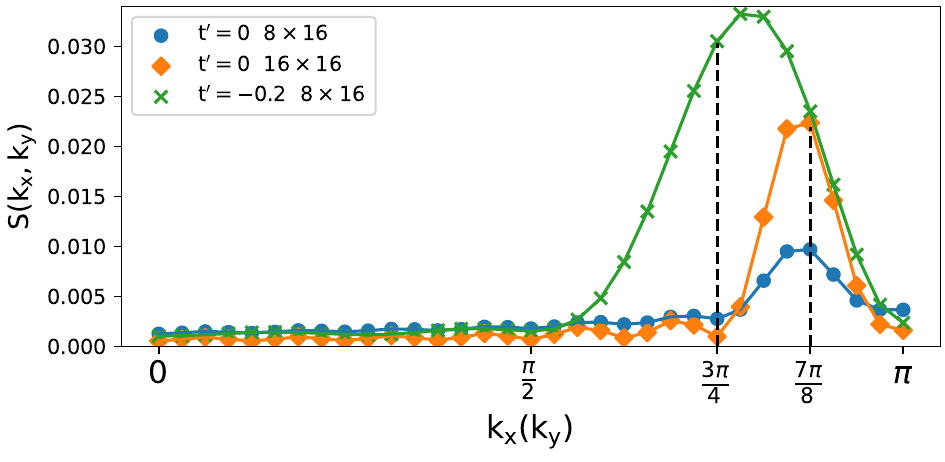}
    \caption{Comparisons of spin structure factors between $t'$=0 and -0.2 for $n$=0.875 and $U$=8 under PBC. All results are from $p$=1 wave-functions, with NN backflow for $t'$=0 and NNN backflow for $t'$=-0.2. Number of k-points are increased from $17^2$ to $65^2$ when calculating spin structure factors. The horizontal axis is $k_x$ for $t'$=-0.2 and $k_y$ for $t'$=0, meanwhile vertical axis is $S(k_x,k_y=\pi)$ for $t'$=-0.2 and $S(k_x=\pi,k_y)$ for $t'$=0.}
    \label{fig:Sk_compare}
\end{figure}

The cases with $t'$=-0.2 are more challenging than those with $t'$=0, as optimizations are more easy to getting trapped in local minima, particularly for larger lattices. We demonstrate that selecting an appropriate initial UHF is crucial for avoiding such local minima and achieving the best final energy~\cite{Hubbard_NN4}. Tab.(\ref{tab:compare_initial_UHF}) compares the energies converged by Tensor-Backflow with initial UHFs corresponding to different values of $U$ on the $12 \times 12$ lattice. For $U$=8, an initial UHF with $U_\mathrm{UHF}$=4 leads to the best converged energy. Ground state patterns for various $U_\mathrm{UHF}$ values on the $12 \times 12$ lattice are presented in App.(~\ref{app:compare_various_U_UHF}).

On the $8 \times 16$ lattice, the ground state pattern for $t'$=-0.2 is shown in Fig.(\ref{fig:Sz_ni_NNN}), with the definitions of the average spin and charge density per site as given in Fig.(\ref{fig:Sz_ni}). For this case, the initial UHF is insufficiently optimized when $U_\mathrm{UHF}$=U. As shown in Fig.(\ref{fig:Sz_ni_NNN}), a width-4 stripe pattern emerges, consistent with the result obtained on the same lattice by NQS~\cite{Hubbard_NN3}. In the NQS optimization, energy comparisons between vertical and horizontal stripe patterns were made, with the horizontal stripe yielding a lower energy~\cite{Hubbard_NN3}. In contrast, in this work, the horizontal stripe naturally emerges after optimization, without the need for pinning-fields.

In Fig.(\ref{fig:Sk_compare}), we compare the spin structure factors for $t'$=0 and $t'$=-0.2. To highlight the differences more clearly, we increase the number of $k$-points from $17^2$ to $65^2$, resulting in a total of 33 $k$-points along the $x$-axis. From the figure, we observe that for $t'$=0, the peaks are sharply defined and centered at $7\pi/8$. However, for $t'$=-0.2, the peak is broader compared to the $t'$=0 case. This broader peak indicates competition between stripe orders with different periods in the ground state for $t'$=-0.2, which makes the optimization more easy to getting trapped in local minima. The increased optimization difficulty for $t'$=-0.2 is further illustrated in Tab.(\ref{tab:compare_initial_UHF}) for the $12 \times 12$ lattice. Therefore, comparisons across various values of $U_\mathrm{UHF}$ will be necessary for $t'$=-0.2 on the $16 \times 16$ lattice in future studies.

\subsection{n=0.8 and n=0.9375}
\label{subsec:n0.8 and 0.9375}
\begin{table}[]
\begin{tabular}{|c|c|c|c|cc|}
\hline
\ \ $n$\ \  & \ \ $U$\ \  & \ \ \begin{tabular}[c]{@{}c@{}}Lattice\\ Size\end{tabular}\ \  & \ \ \begin{tabular}[c]{@{}c@{}}Backflow\\ Type\end{tabular}\ \  & \multicolumn{1}{c|}{\ \ $E_{p=0}$\ \ \ } & \ \ \ $E_{p=1}$\ \  \\ \hline
\multirow{4}{*}{0.8}    & 4                  & $10\times 20$ & \multirow{3}{*}{NN} & -1.1012 & -1.1037 \\ \cline{2-3}
                        & \multirow{3}{*}{8} & $10\times 20$ &                     & -0.8755 & -0.8793 \\ \cline{3-3}
                        &                    & $10\times 10$ &                     & -0.8801 & -0.8848 \\ \cline{3-4}
                        &                    & $10\times 10$ & all-site            & -0.8809 & -0.8853 \\ \cline{1-4}
\multirow{3}{*}{0.9375} & \multirow{3}{*}{4} & $8\times 16$  & \multirow{3}{*}{NN} & -0.9384 & -0.9409 \\ \cline{3-3}
                        &                    & $8\times 24$  &                     & -0.9394 & -0.9421 \\ \cline{3-3}
                        &                    & $8\times 32$  &                     & -0.9401 & -0.9426 \\ \hline
\end{tabular}
	\caption{Energy results for the filling $n$=0.8 and $n$=0.9375, with $t'$=0 under PBC. For all cases in the table, initial UHFs are insufficiently optimized with $U_\mathrm{UHF}=U$. On the $10\times 10$ lattice, all-site backflow has moderate improvement comparing to NN backflow for $n$=0.8 and $U$=8. }
	\label{tab:energies_n0.8_0.9375}
\end{table}
\begin{figure}[t]
\includegraphics[width=\columnwidth]{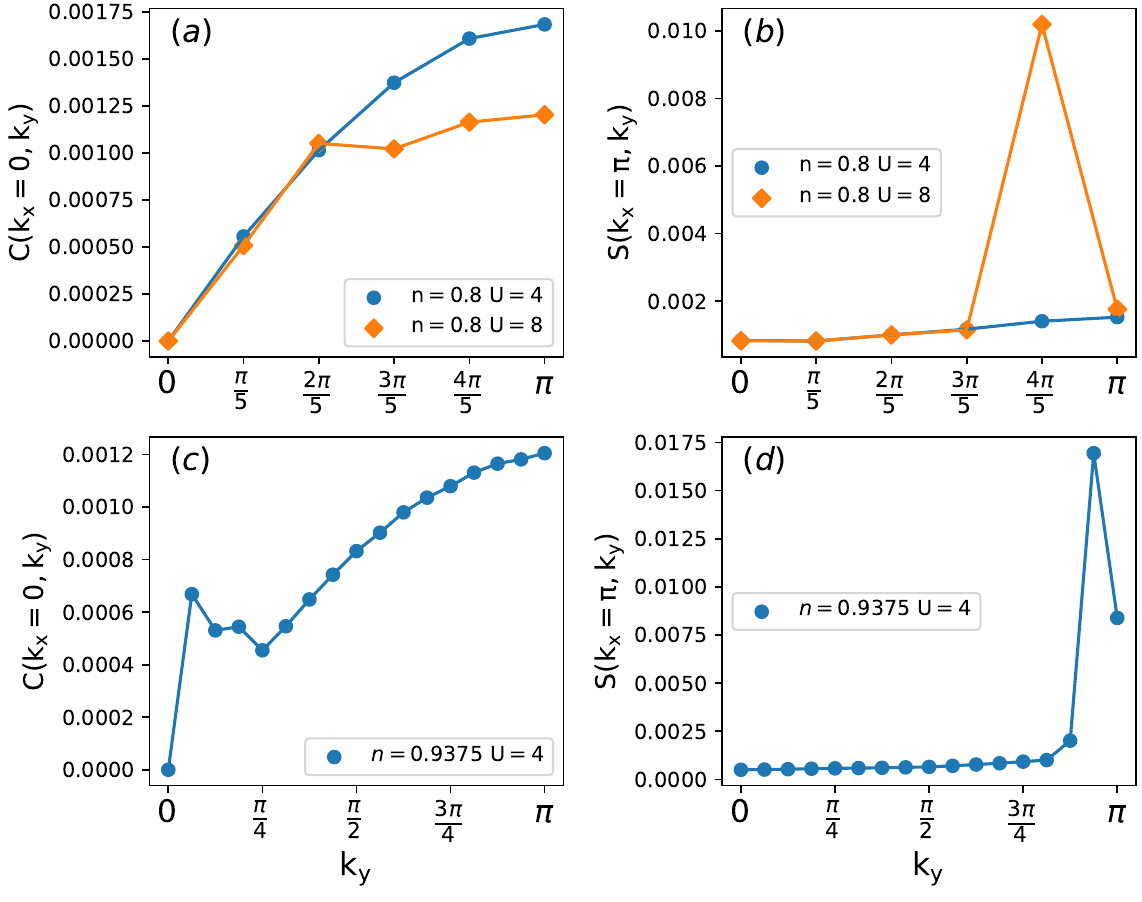}
\caption{Charge and spin structure factors on the $10\times 20$ lattice for $n$=0.8 and $t'$=0. All results are from $p$=1 wave-functions with NN backflow in wave-functions. (a)(b) and on the $8\times 32$ lattice for $n$=0.9375 (c)(d), under PBC. (a)(b)For $n$=0.8, the spin structure factor peaks at $4\pi/5$ when $U$=8, however there are no obvious peaks when $U$=4. For both $U$=4,8, there are no obvious peaks in the charge structure factor. (c)(d)For $n$=0.9375, the spin structure factor peaks at $15\pi/16$ when $U$=4, meanwhile there are no obvious peaks in the charge structure factor. }
\label{fig:Sk_Ck_n0.8_0.9375}
\end{figure}

For other fillings, such as $n$=0.8 and $n$=0.9375, energy results are provided in Tab.(\ref{tab:energies_n0.8_0.9375}). In the case of $n$=0.8 and $U$=8 on the $10\times 10$ lattice, the improvement from all-site backflow over NN backflow is relatively moderate. Therefore, for a Hamiltonian with NN hoppings, the primary contribution to the state representation comes from the NN backflow.

Fig.(\ref{fig:Sk_Ck_n0.8_0.9375}) shows the charge and spin structure factors for $n$=0.8 on the $10\times 20$ lattice and for $n$=0.9375 on the $8\times 32$ lattice. According to the phase diagram from AFQMC~\cite{AFQMC_phase_diagram}, the critical interaction strength $U_c$ that separates the disordered and ordered ground states increases with the doping level $h$. For $n$=0.8, the spin structure factor for $U$=8 peaks at $4\pi/5$, whereas for $U$=4, no obvious peak is observed. In the charge structure factor, an increasing peak appears as $U$ increases, especially at $U$=8. For $n$=0.9375, the spin structure factor peaks at $15\pi/16$, but no clear peak is seen in the charge structure factor. The results for both $n$=0.8 and $n$=0.9375 in Fig.(\ref{fig:Sk_Ck_n0.8_0.9375}) are consistent with the phase diagram from AFQMC~\cite{AFQMC_phase_diagram,AFQMC_n0.9375}.

\section{Discussions and conclusions}
\label{sec:conclusions}
In this work, we benchmark the Tensor-Backflow method, followed by a Lanczos step, across a range of fillings $n$, interaction strengths $U$, NNN hopping strengths $t'$, lattice sizes, and boundary conditions. Compared to state-of-the-art methods such as fPEPS~\cite{Hubbard_fPEPS} and NQS~\cite{Hubbard_NN3,Hubbard_NN4}, the energies achieved by Tensor-Backflow are competitive or even lower. In addition to energy results, we investigate the ground state properties, including the spin ($s_i^z$) and density ($n_i$) patterns, as well as the charge and spin structure factors. For $t'$=0, the state properties obtained by Tensor-Backflow are consistent with those from fPEPS~\cite{Hubbard_fPEPS} and NQS~\cite{Hubbard_NN3} for $n$=0.875, and with AFQMC for $n$=0.8~\cite{AFQMC_phase_diagram} and $n$=0.9375~\cite{AFQMC_n0.9375}. For $t'$=-0.2 on the $8\times 16$ lattice, a width-4 horizontal stripe is obtained, consistent with the result from NQS~\cite{Hubbard_NN3}. On the $12\times 12$ lattice, the achieved energy is lower than the best-known NQS result~\cite{Hubbard_NN3}, with a width-4 stripe identified based on the charge density $n_i$.

\begin{figure}[t]
\includegraphics[width=0.8\columnwidth]{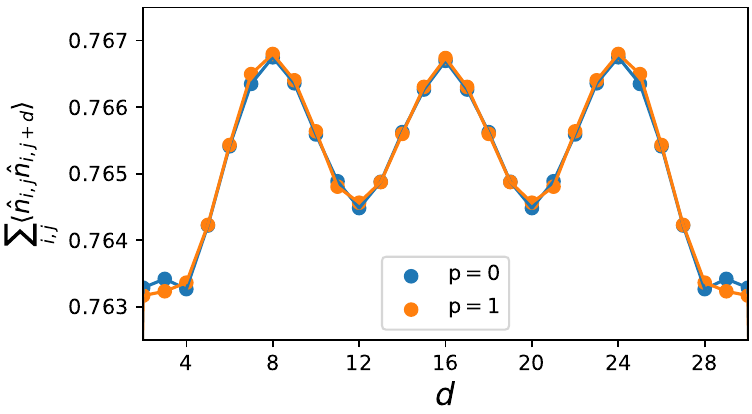}
\caption{Density-density correlations for $n$=0.875, $U$=8 and $t'$=0 on the $8\times 32$ lattice under PBC. With NN backflow in the wave-function. Blue dots depict the $p$=0 wave-function and yellow dots depict the $p$=1 wave-function. For both $p$=0 and $p$=1, results are obtained based on 46080 MC samples, with the interval between two samples twice as the lattice size. }
\label{fig:ninj_n0.875_U8_8x32}
\end{figure}

Our results show that, although a Lanczos step significantly improves the energy, the properties of the state are determined primarily by the $p$=0 wave-function. Specifically, a Lanczos step does not change the fundamental properties of the original wave-function, as the patterns of $s_i^z$ and $n_i$ remain similar between the $p$=0 and $p$=1 wave-functions. For instance, Fig.(\ref{fig:ninj_n0.875_U8_8x32}) shows the density-density correlations from both the $p$=0 and $p$=1 wave-functions for $n$=0.875 and $U=8$ on the $8\times 32$ lattice under PBC. As seen in the figure, the CDW with a period of 8 is already present in the $p$=0 wave-function, and no significant differences are observed in the correlations for the $p$=1 wave-function. Therefore, the differences between the two-body correlations from the $p$=0 and $p$=1 wave-functions are negligible.

\begin{figure}[t]
\includegraphics[width=\columnwidth]{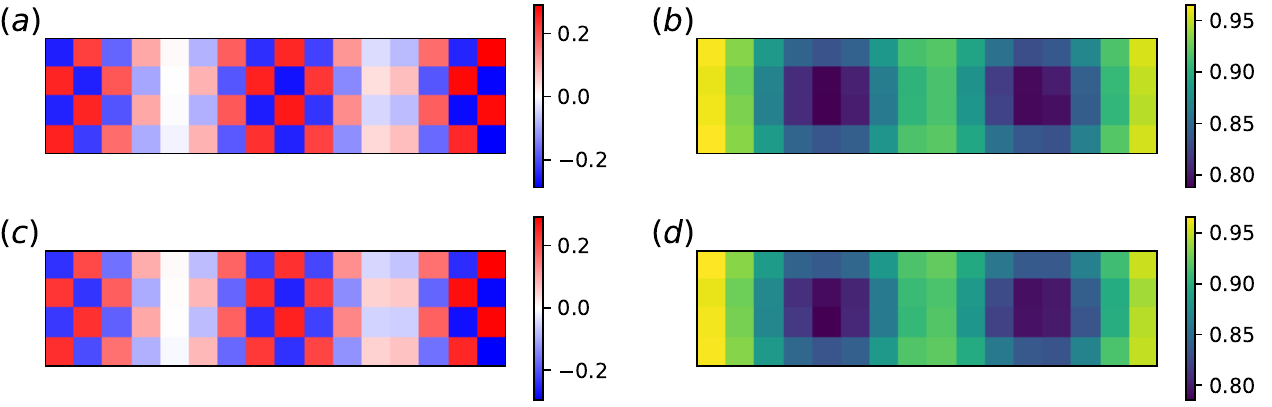}
\caption{Comparisons of ground state patterns between two optimization strategies in Sec.~(\ref{subsec:optimization details}). The case is $n$=0.875, $U$=8 and $t'$=0 on the $4\times 16$ lattie under OBC, with NN backflow in wave-functions. Left-sided figures denote the average spin per site: $s_i^z$, and right-sided figures denote the average density per site: $n_i$. (a)(b) and (c)(d) are achieved by optimization strategies 1 and 2, respectively. }
\label{fig:optimization_compare}
\end{figure}

Pair correlations are essential for determining the presence of superconductivity~\cite{tJ_fPEPS, pair_pair_correlation_1, pair_pair_correlation_2}. In this work, to mitigate the computational complexity, we evaluate $d$-wave pair correlations using the $p$=0 wave-functions, based on the definition of pair correlations in the literature~\cite{tJ_fPEPS}. For the case of $n$=0.875 and $U$=8 on the $16 \times 16$ lattice under PBC, we find that pair correlations are on the order of $\mathcal{O}(10^{-5})$ when the distance between pairs is $L/2$, where $L$ is the side length of the square lattice, for both $t'$=0 and $t'$=-0.2. Since optimizations are easy to getting trapped in local minima, particularly for cases with $t'$=-0.2, the future work includes comparing energies with various stripe orders by applying pinning-fields and calculating pair correlations using $p$=1 wave-functions.

Based on the results presented in this work, Tensor-Backflow provides an efficient state representation for the lattice Fermi-Hubbard model. There are several methods for improving energy precision. For instance, compare energies for different initializations or investigate ground state patterns both with and without the application of pinning fields.
At present, the wave-function is not symmetry-enforced, meaning translational invariance is not preserved during optimization. For systems with PBC, energy precision could be further enhanced by enforcing translational invariance in the wave-function.
Additionally, the Tensor-Backflow method is flexible with respect to lattice shapes and boundary conditions, making it a potentially efficient and universal approach for solving lattice fermion systems.

\section{Acknowledgement}
X. Liang thanks useful discussions with S. Zhang, W.-Y. Liu and A. Chen.
X. Liang thanks Eric J. Walter for resource allocations on the William \& Mary Sciclone cluster.
The author acknowledges William \& Mary Research Computing and the Texas Advanced Computing Center (TACC) at The University of Texas at Austin for providing computational resources that have contributed to the results reported within this paper.
Part of numerical results were obtained on the Bridges-2 at Pittsburgh Supercomputing Center through allocation (phy240244).

\begin{appendices}
\section{Comparing ground state patterns of two optimization strategies}
\label{app:optimization_compare}
\begin{figure}[t]
\includegraphics[width=\columnwidth]{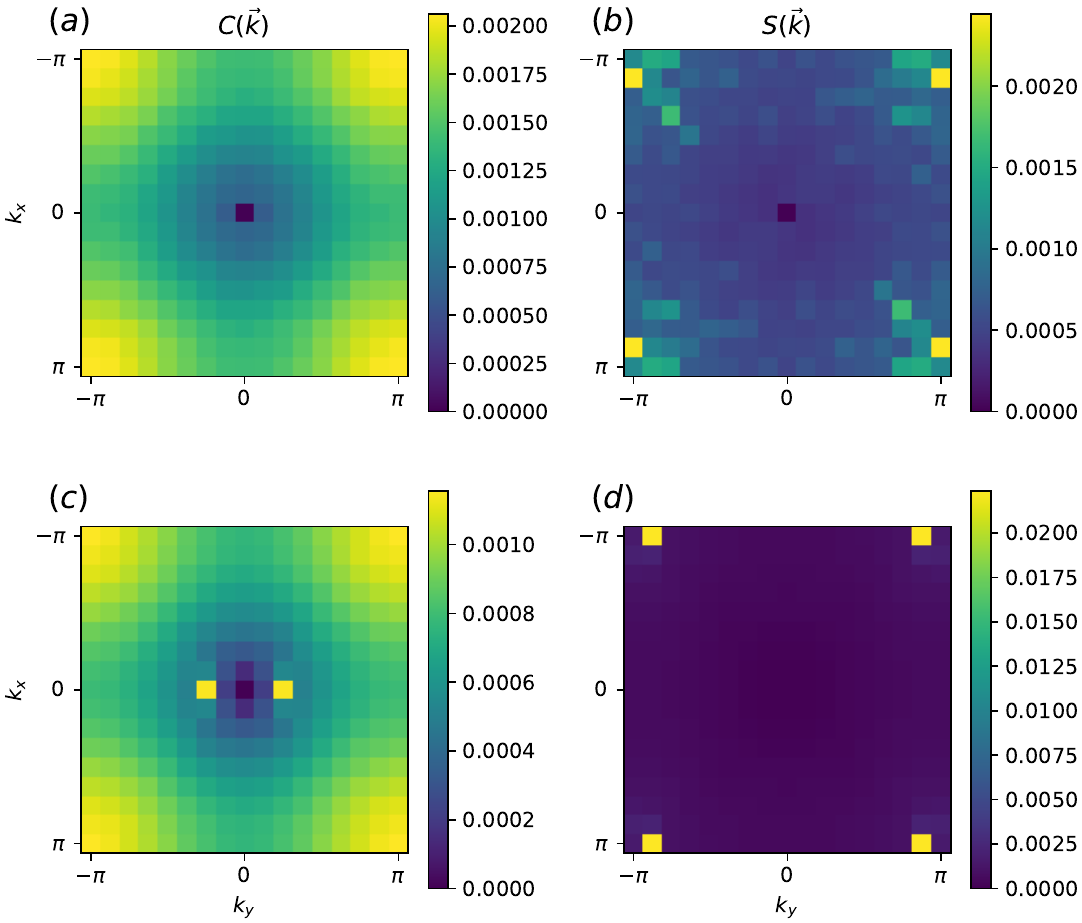}
\caption{Comparisons of structure factors $C(\mathbf{k})$ and $S(\mathbf{k})$ from (a)(b)the initial UHF and (c)(d)the converged wave-function. The case is $n$=0.875, $U$=8 and $t'$=0 on the $16\times 16$ lattice under PBC, with NN backflow in the wave-function. }
\label{fig:UHF_BW_compare_1}
\end{figure}
\begin{figure}[t]
\includegraphics[width=\linewidth]{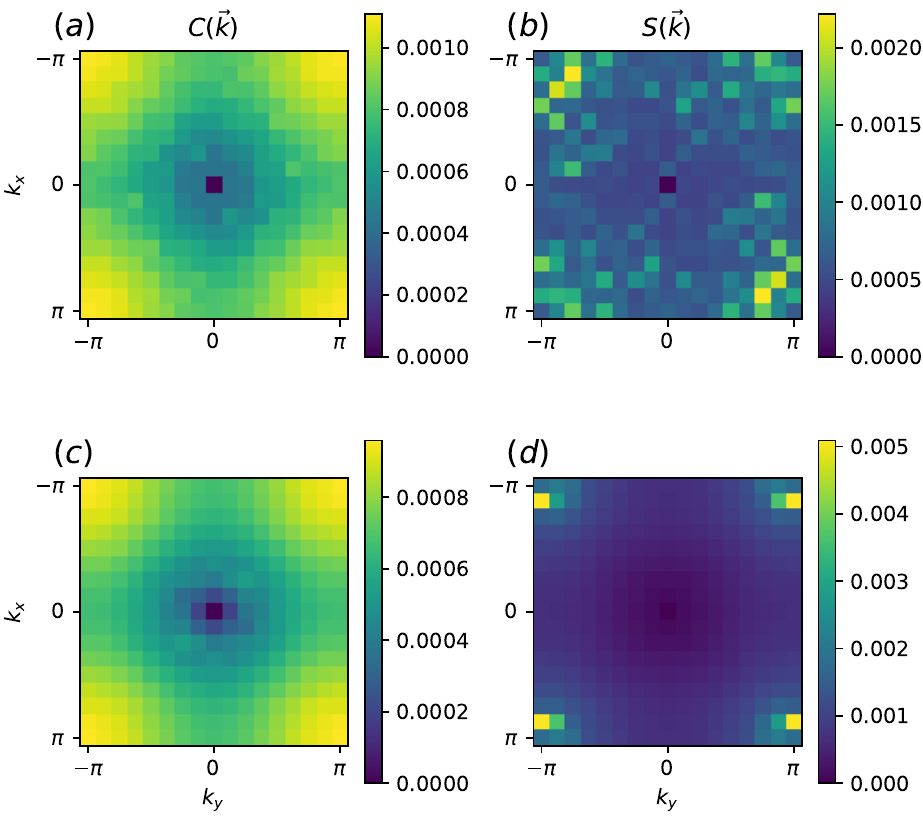}
\caption{Comparisons of structure factors $C(\mathbf{k})$ and $S(\mathbf{k})$ from (a)(b)the initial UHF, (c)(d)the converged wave-function. The case is $n$=0.875, $U$=10 and $t'$=0 on the $16\times 16$ lattice under PBC, with NN backflow in the wave-function. }
\label{fig:UHF_BW_compare_2}
\end{figure}
\begin{figure}[t]
\includegraphics[width=\linewidth]{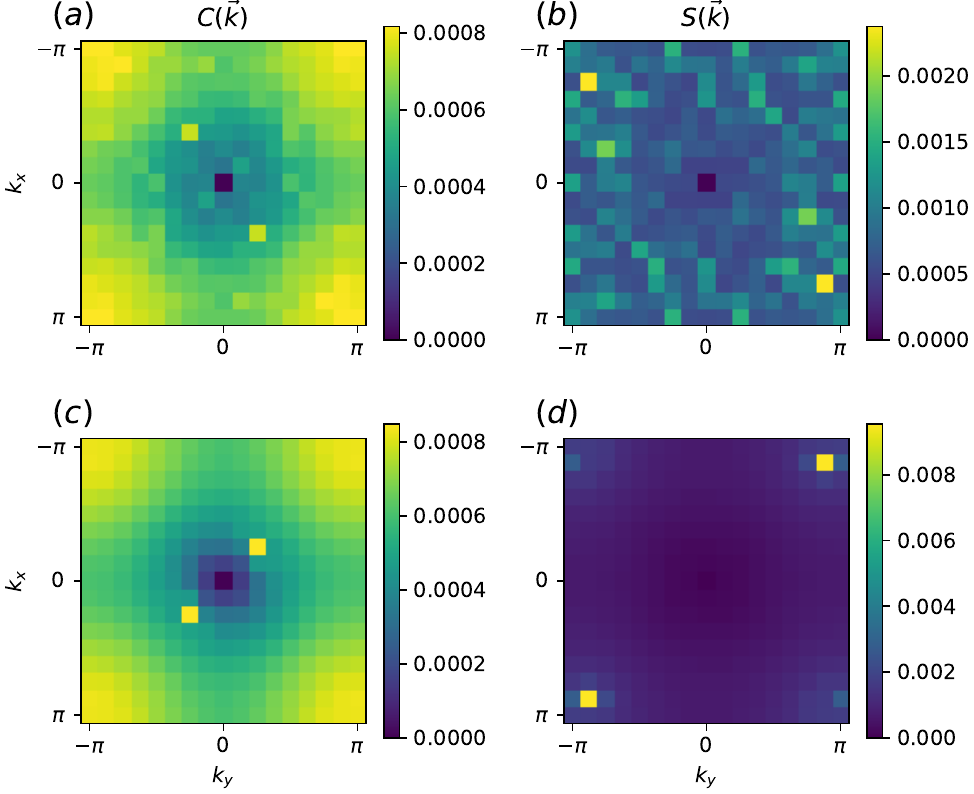}
\caption{Comparisons of structure factors $C(\mathbf{k})$ and $S(\mathbf{k})$ from (a)(b)the initial UHF, (c)(d)the converged wave-function. The case is $n$=0.875, $U$=12 and $t'$=0 on the $16\times 16$ lattice under PBC, with NN backflow in the wave-function.}
\label{fig:UHF_BW_compare_3}
\end{figure}
   \begin{figure}[t]
\includegraphics[width=0.9\linewidth]{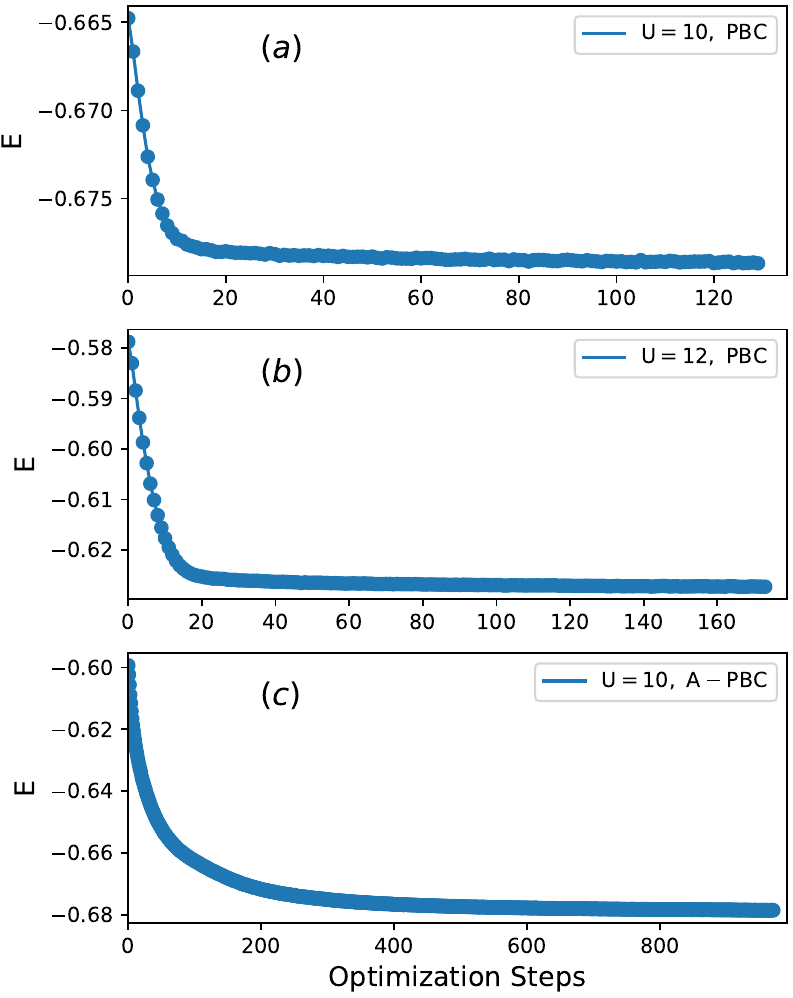}
\caption{Energy convergence for $n$=0.875 and $t'$=0 on the $16\times 16$ lattice when the wave-function is transferred from $U$=8, with NN backflow in wave-functions. More steps are required when the case has larger difference than $U$=8 and PBC. Converged energies are -0.6787, -0.6272 and -0.6786 in (a)(b) and (c), respectively.}
\label{fig:transfer_convergence}
\end{figure}
\begin{figure}[t]
\includegraphics[width=0.85\linewidth]{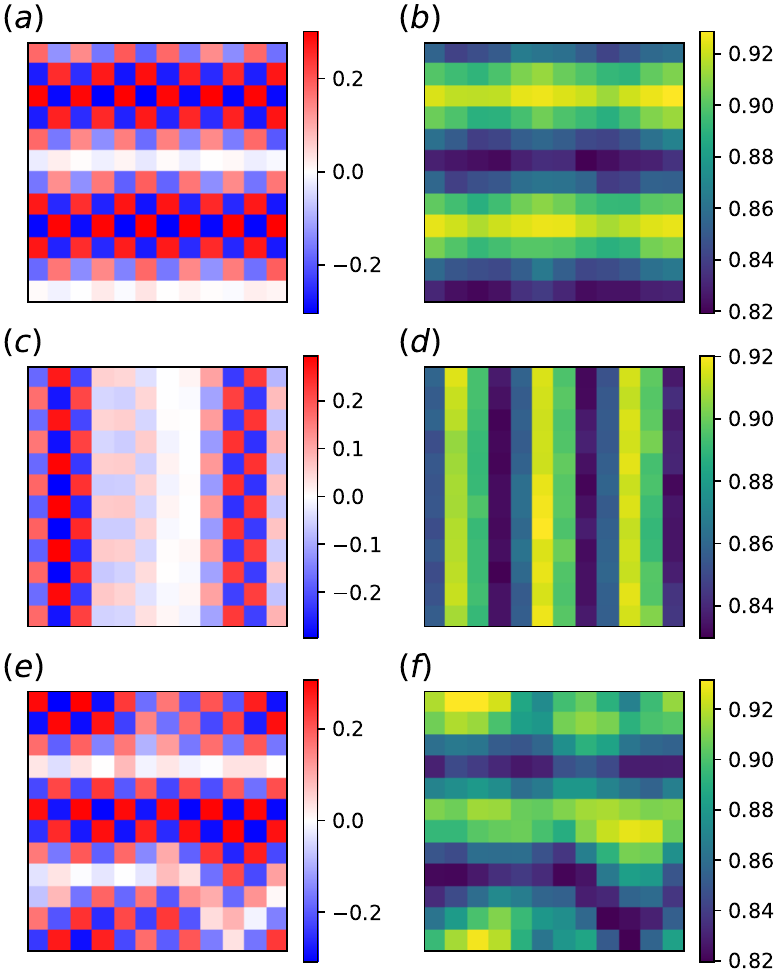}
\caption{Ground state patterns for $n$=0.875, $U$=8, $t'$=-0.2 on the $12\times 12$ lattice under PBC for various $U_\mathrm{UHF}$, with NNN backflow in wave-functions. Left-sided figures denote the averate spin per site: $s_i^z$ and right-sided figures denote the averate density per per site: $n_i$. Results of $U_\mathrm{UHF}$=0, 4, 8 are depicted in subfigures (a)(b), (c)(d) and (e)(f), respectively. All results are from $p$=1 wave-functions. No pinning-fields are applied during optimizations. }
\label{fig:compare_various_U_UHF}
\end{figure}

Fig.(\ref{fig:optimization_compare}) compares the ground state patterns obtained from the two optimization strategies described in Sec.~(\ref{subsec:optimization details}), using $p$ = 0 wave-functions. The case corresponds to $n$=0.875, $U$=8 on the $4 \times 16$ lattice under OBC. From the figure, both optimization strategies lead to similar ground state patterns. Moreover, the patterns in Fig.(\ref{fig:optimization_compare}) are in agreement with those from state-of-the-art NQS results~\cite{Hubbard_NN3}.

\section{Comparisons between initial UHFs and subsequent Tensor-Backflow wave-functions}
\label{app:compare_UHF_backflow}

We demonstrate that the Tensor-Backflow method produces a state that is fundamentally different from the initial UHF, as revealed by the charge and spin structure factors $C(\mathbf{k})$ and $S(\mathbf{k})$. For $n$=0.875 on the $16 \times 16$ lattice under PBC, we show results for $U$=8, 10, and 12 in Fig.(\ref{fig:UHF_BW_compare_1}), Fig.(\ref{fig:UHF_BW_compare_2}), and Fig.(\ref{fig:UHF_BW_compare_3}), respectively. As described in Sec.~(\ref{subsec:n0.875}), the correlations in both $C(\mathbf{k})$ and $S(\mathbf{k})$ are classical in nature, and are therefore computed by averaging over MC samples. Specifically, each structure factor is calculated using 46,080 MC samples, with the results for the UHF and Tensor-Backflow states derived from the $p$=0 and $p$=1 wave-functions, respectively.

In our calculations, each MC sample is flattened from two-dimensional to one-dimensional using the same procedure, and each structure factor is computed consistently from the MC samples. By comparing panels (a) and (b) with (c) and (d) in Fig.(\ref{fig:UHF_BW_compare_1}), Fig.(\ref{fig:UHF_BW_compare_2}), and Fig.(\ref{fig:UHF_BW_compare_3}), we observe that the properties of the converged wave-function are not simply inherited from the initial UHF. Moreover, the Tensor-Backflow method allows the wave-function to override the properties of the initial UHF. In all cases, we set $U_\mathrm{UHF}=U$, and the initial UHFs are not fully optimized in any of the three scenarios.

\section{Energy convergence of transferring from $U$=8}
\label{app:transfer_convergence}

Fig.(\ref{fig:transfer_convergence}) shows the energy convergence when transferring the converged wave-function from $U$=8 and PBC to other cases. Due to the quality of the initial state, the number of optimization steps required for transfer is significantly smaller compared to direct optimization.
For the cases labeled (a), (b), and (c) in Fig.(\ref{fig:transfer_convergence}), the energies obtained by direct optimization are -0.6771, -0.6264, and -0.6771, respectively. In contrast, the energies achieved by transferring the wave-function from $U$=8 are -0.6787, -0.6272, and -0.6786, respectively.

\section{Comparing ground state patterns for various $U_\mathrm{UHF}$ for $t'$=-0.2}
\label{app:compare_various_U_UHF}

For the case of $n$=0.875, $U$=8, $t'$=-0.2 on the $12\times 12$ lattice under PBC, comparisons of ground state patterns of $U_\mathrm{UHF}$=0, 4 and 8 are depicted in Fig.(\ref{fig:compare_various_U_UHF}), with converged energies from Tab.(\ref{tab:compare_initial_UHF}).
All optimization are performed without pinning-fields applied.

From the figure, different periods of stripes are achieved with different $U_\mathrm{UHF}$~\cite{Hubbard_NN4}. Therefore, $U_\mathrm{UHF}$ is essential to avoid local minima when optimizing Tensor-Backflow. Based on the density $n_i$, the lowest energy of -0.7420 is achieved by the width-4 stripe, yielding to the state-of-the-art NQS result~\cite{Hubbard_NN3}.

\end{appendices}

\end{document}